\documentclass[%
 reprint,
 superscriptaddress,
 showpacs,preprintnumbers,
 amsmath,amssymb,
 aps,
 prb,
]{revtex4-1}

\usepackage{graphicx}
\usepackage{dcolumn}
\usepackage{bm}

\usepackage{upgreek}
\usepackage{amsmath}

\begin{document}


\title{Engineering light absorption at critical coupling via bound states in the continuum}

\author{Shuyuan Xiao}
\email{syxiao@ncu.edu.cn}
\affiliation{Institute for Advanced Study, Nanchang University, Nanchang 330031, China}
\affiliation{Jiangxi Key Laboratory for Microscale Interdisciplinary Study, Nanchang University, Nanchang 330031, China}

\author{Xing Wang}
\author{Junyi Duan}
\affiliation{Institute for Advanced Study, Nanchang University, Nanchang 330031, China}

\author{Chaobiao Zhou}
\author{Xiaoying Qu}
\affiliation{College of Mechanical and Electronic Engineering, Guizhou Minzu University, Guiyang 550025, China}

\author{Tingting Liu}
\email{ttliu@hue.edu.cn}
\affiliation{School of Physics and Electronics Information, Hubei University of Education, Wuhan 430205, China}

\author{Tianbao Yu}
\email{yutianbao@ncu.edu.cn}
\affiliation{Institute for Advanced Study, Nanchang University, Nanchang 330031, China}
\affiliation{Jiangxi Key Laboratory for Microscale Interdisciplinary Study, Nanchang University, Nanchang 330031, China}
\affiliation{Department of physics, Nanchang University, Nanchang 330031, China}

\begin{abstract}
Recent progress in nanophotonics is driven by the desire to engineer light-matter interaction in two-dimensional (2D) materials using high-quality resonances in plasmonic and dielectric structures. Here, we demonstrate a link between the radiation control at critical coupling and the metasurface-based bound states in the continuum (BIC) physics, and develop a generalized theory to engineer light absorption of 2D materials in coupling resonance metasurfaces. In a typical example of hybrid graphene-dielectric metasurfaces, we present the manipulation of absorption bandwidth by more than one order of magnitude by simultaneously adjusting the asymmetry parameter of silicon resonators governed by BIC and the graphene surface conductivity while the absorption efficiency maintains maximum. This work reveals the generalized role of BIC in the radiation control at critical coupling and provides promising strategies in engineering light absorption of 2D materials for high-efficiency optoelectronics device applications, e.g., light emission, detection and modulation.
\end{abstract}

\maketitle


\section{\label{sec1}Introduction}
Over the past decade, two-dimensional (2D) materials have emerged as promising core elements in the next generation optoelectronic devices due to their exceptional optical and electric properties\cite{Xia2014}. The main obstacle to practical applications is the relatively weak light-matter interaction, and it is of paramount importance to enhance light absorption in these atomically thin films. J. R. Piper introduced a concept of critical coupling to engineer light absorption with a photonic crystal guided resonance. The absorption in the top graphene can reach maximum when the external radiation rate of the guided resonance equals to the internal dissipative loss rate of graphene\cite{Piper2014, Liu2014}. Such a system shows great advantages over Fabry−Perot cavity in the simplicity in fabrication, robust performance and compatible features, and afterwards inspires the exploration of absorption in the whole family of 2D materials\cite{Guo2016, Jiang2017, Li2017, Jiang2018, Qing2018, Wang2019, Liu2019, Cheng2020}.

In above-mentioned critical coupling systems, several key figures of the absorption properties including the central location, efficiency and bandwidth can be tuned by adjusting the geometric parameters of photonic crystal slabs. The absorption engineering essentially originates from the control of the radiation rate of the guided resonance. In reality, such kind of radiation control based on the dependence of geometrical parameters is ubiquitous in the field of nanophotonics, which is definitely not limited to the guided resonance in photonic crystal slabs, but also universally applicable to resonance modes in a variety of seemingly different photonic structures such as trapped mode\cite{Fedotov2007, Zhang2013, Sayanskiy2019}, toroidal mode\cite{Savinov2014, Tuz2018, Fan2018, He2018}, non-radiating anapole mode\cite{Miroshnichenko2015, Lukextquotesingleyanchuk2017, Yang2018} and supercavity mode\cite{Kodigala2017, Rybin2017} that are fundamentally linked to the physics of bound states in the continuum (BIC). 

In this work, we demonstrate a link between the radiation control at critical coupling and the fascinating BIC physics, and develop a general theory to engineer light absorption of 2D materials in coupling resonance metasurfaces. As an example, we present to place a monolayer graphene on top of asymmetric pairs of silicon nanobars. The maximum absorption of graphene is attained under the critical coupling condition when the radiation rate of the trapped mode in the resonators is equal to the dissipative loss of graphene. By simultaneously adjusting the asymmetry parameter for the resonators and the graphene surface conductivity, the absorption bandwidth can be flexibly tuned in good agreement with the theoretical predictions while the absorption efficiency maintains maximum. This work reveals the generalized role of BIC in the radiation control at critical coupling, and delineates the in-depth conceptual framework and novel promising strategies in engineering light absorption of 2D materials for smart design of compact optoelectronic devices.

\section{\label{sec2}Theoretical Formalism}
The coupled mode theory (CMT) is utilized to provide insights into the input and output properties of a resonance system. We consider the behavior of a mirror-symmetric resonator that possesses a single resonance mode coupling with outside circumstances via two identical ports. For a lossless system, the dynamic equations can be written as\cite{Haus1984, Fan2003}
\begin{eqnarray}
\frac{da}{dt}&=&(i\omega_{0}-\gamma)a+D^{T}|s_{+}\rangle,\label{eq1} \\
|s_{-}\rangle&=&C|s_{+}\rangle+Da,\label{eq2}
\end{eqnarray}
where $a$ is the resonance amplitude, $\omega_{0}$ and $\gamma$ represent the resonance frequency and radiation rate, $|s_{+}\rangle=[s_{1+},s_{2+}]^{T}$, $|s_{-}\rangle=[s_{1-},s_{2-}]^{T}$ and $D$ represent the input waves, the output wave and the coupling matrix, respectively, and $C$ is the scattering matrix of the direct coupling pathway. In such system, the resonance mode is excited by the input waves $|s_{+}\rangle$ from ports 1, 2 with the coupling matrix $D^{T}$, and the excited resonance mode couples with the output waves $|s_{-}\rangle$ at the ports with the coupling matrix $D$. The dynamic equations are constrained by energy-conservation and time-reversal symmetry considerations. For the case where the input wave with unit amplitude is incident from only a single port, the stored energy in the resonator can be derived from the general formalism Eq. (\ref{eq1}) and (\ref{eq2})
\begin{eqnarray}
|a|^{2}=\frac{\gamma}{(\omega-\omega_{0})^2+\gamma^{2}}.\label{eq3}
\end{eqnarray}

Further, when the dissipative loss is inserted into the system and the underlying symmetry of the structure remains unchanged, the dynamic equation Eq. (\ref{eq1}) can be accordingly modified by adding the effect of the dissipative loss rate $\delta$, i.e., $da/dt=(i\omega_{0}-\gamma-\delta)a+D^{T}|s_{+}\rangle$. Then $|a|^{2}$ in Eq. (\ref{eq3}) is amended by replacing the term $\gamma^{2}$ in the denominator with $(\gamma+\delta)^{2}$, and the absorption in the resonance system can be expressed as
\begin{eqnarray}
A=\frac{2\delta\gamma}{(\omega-\omega_{0})^2+(\gamma+\delta)^{2}}.\label{eq4}
\end{eqnarray}
As can be seen in Eq. (\ref{eq4}), the absorption of the resonance system will exhibit a symmetric Lorentzian lineshape, and three key paramaters, $\omega_{0}$, $\gamma$ and $\delta$ determine the absorption characteristics. More specifically, the resonance frequency $\omega_{0}$ predicts the central location of the absorption spectrum, i.e., the center frequency of the Lorentzian lineshape, and the radiation rate $\gamma$ and the dissipative loss rate $\delta$ jointly defines the efficiency and bandwidth of absorption. At the resonance frequency $\omega=\omega_{0}$, the absorption $A_{0}$ is determined by the ratio between $\gamma$ and $\delta$
\begin{eqnarray}
A_{0}=\frac{2}{\frac{\gamma}{\delta}+\frac{\delta}{\gamma}+2},\label{eq5}
\end{eqnarray} 
which will attain a theoretical maximum of 0.5 when $\gamma=\delta$. This is the so-called critical coupling condition for a single resonance in a two-port system. 

Meanwhile, the absorption bandwidth is usually defined as the full width at half maximum (FWHM). Assuming the half maximum absorption $A_{1}=A_{0}/2$ occurs at $\omega_{1}$, $\Gamma^{\text{FWHM}}=2|\omega_{1}-\omega_{0}|$ can be calculated as
\begin{eqnarray}
\Gamma^{\text{FWHM}}=2(\gamma+\delta).\label{eq6}
\end{eqnarray}

In association with the general wave phenomenon of BIC, the radiaiton rate $\gamma$ of a resonance mode is considered as mode inverse radiation lifetime. An ideal symmetry-protected BIC has no access to radiation channels and thus can be treated as a resonance with infinite lifetime and vanishing linewidth, i.e., $\gamma=0$. Then the quasi-BICs concept is introduced in practice that couples to the extended waves and radiates as a leaky resonance with $\gamma>0$ when the symmetry of a resonator is broken to generate radiation channels. Formulated from the analytical calculations, the radiation rate $\gamma$ is calculated by the sum of radiation losses into all the radiation channels\cite{Koshelev2018, Li2019}
\begin{eqnarray}
\gamma=c\sum_{i=x,y}|D_{i}|^{2},\label{eq7}
\end{eqnarray}
with the coupling amplitudes $D_{i}$
\begin{eqnarray}
D_{x,y}=-\frac{k_{0}}{\sqrt{2S_{0}}}(p_{x,y}\mp\frac{1}{c}m_{y,x}+\frac{ik_{0}}{6}Q_{xz,yz}),\label{eq8}
\end{eqnarray}
where $k_{0}$ is the incident wave vector in the free space, $S_{0}$ is the unit cell area, and $p_{\alpha}$, $m_{\alpha}$ and $Q_{\alpha\beta}$ ($\alpha,\beta=x,y,z$) are the components of electric dipole, magnetic dipole and electric quadrupole moments in the irreducible representations, respectively.   

Considering a situation where the in-plane symmetry along the $x$ axis is broken, it follows that $D_{y}=0$ since the electric field components $E_{x}$ and $E_{y}$ are classified as even and odd functions with respect to this symmetry. Also, $m_{y}$ and $Q_{xz}$ are both equal to zero because of the symmetry $E_{x}(-z)=E_{x}(z)$. Then a simplified analytical expression can be written as 
\begin{eqnarray}
\gamma=\frac{k_{0}^{2}c}{2S_{0}}|p_{x}|^{2}.\label{eq9}
\end{eqnarray}
$p_{x}=\pm\alpha p_{0}$ is defined as the net dipole moment in an asymmetric resonator, where $p_0$ is the electric dipole moment in an according symmetric case and $\alpha$ is the asymmetry parameter. Then for a general kind of metasurfaces with asymmetric unit cell, the radiation rate $\gamma$ of the quasi-BIC can be found to be dependent on the asymmetry parameter
\begin{eqnarray}
\gamma=\frac{k_{0}^{2}c}{2S_{0}}|p_{0}|^{2}\alpha^{2}.\label{eq10}
\end{eqnarray}

Substituing the expression of $\gamma$ in Eq. (\ref{eq10}) into the absorption bandwidth at critical couping in Eq. (\ref{eq6}), the direct dependence of FWHM on the asymmetry parameter $\alpha$ can be finally obtained
\begin{eqnarray}
\Gamma^{\text{FWHM}}=\frac{2k_{0}^{2}c}{S_{0}}|p_{0}|^{2}\alpha^{2},\label{eq11}
\end{eqnarray}
i.e., $\Gamma^{\text{FWHM}}\propto\alpha^{2}$, following the universal rules for the asymmetric metasurfaces with high-quality resonance driven by BIC. Based on the quadratic scalability of FWHM, a simple but rigorous way to engineer light absorption at critical coupling is suggested by adjusting the asymmetry parameter for the quasi-BIC resonators. Consequently, a general physical link between the radiation control at critical coupling and the BIC physics is theoretically established.

\section{\label{sec3}Results and Discussions}
\begin{figure}[htbp]
\centering
\includegraphics
[scale=0.32]{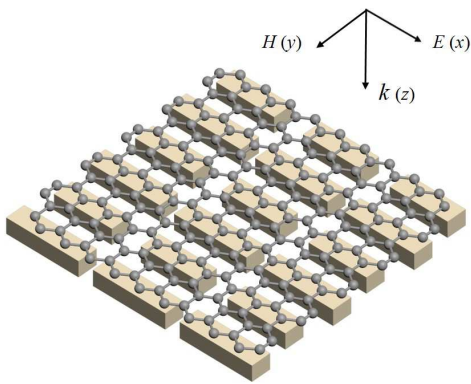}
\caption{\label{fig1} Schematic of the proposed hybrid graphene-dielectric metasurfaces.}
\end{figure}

We focus on one of the examples in which a common 2D material of monolayer graphene is deposited on top of a typical dielectric metasurfaces that supports magnetic dipole mode, as shown in FIG. \ref{fig1}. The unit cell is periodically arranged with a lattice constant $P=900$ nm, and composed of a pair of parallel, geometrically asymmetric nanobars with a fixed separation of $D=250$ nm. The thickness and width of the nanobars are the same value of $H=160$ nm and $W=200$ nm, while the lengths of them are set with a variable $L_{1}$ and a fixed $L_{2}=750$ nm for each case that is investigated. In accordance with the fact that a symmetry-protected BIC can be transformed to a quasi-BIC by simply breaking the symmetry of the system, it is predictable that the quasi-BIC state can be realized by opening a radiation channel via introducing a perturbation in the nanobar pair.

Silicon is chosen as the dielectric building material because it is characterized by high refractive index and negligible absorption loss in the studied spectral range of near infrared. For simplification, silicon is assumed to be lossless with a refractive index of $n=3.5$. On the other hand, the monolayer graphene on the top can be considered as an ultra-thin lossy film. The optical properties of graphene can be captured by random phase approximation (RPA) in the local limit, where the surface conductivity is expressed by the sum of intraband and interband contributions\cite{Zhang2015, Xiao2016}. In particular, it is capable to be controlled as the Fermi level varies. As a consequence of increasing the Fermi level, the imaginary part shows a continuous increase and the real part reduces noticeably at the critical point of Fermi level exceeding the Dirac point by half photon energy, i.e., $E_{\text{F}}>\hbar\omega/2$ due to the Pauli-blocked effect. The tunable surface conductivity of graphene enables the flexible control of absorption in the resonance system. The numerical simulations are conducted using the finite-difference time-domain (FDTD) method. In the calculations, the moderate mesh grid is adopted to make good trade-off between accuracy, memory requirements and simulation time. The linearly polarized plane wave is incident along the $-z$ direction, and the periodic boundary conditions are utilized in the $x$ and $y$ directions and the perfect matching layers are adopted in the $z$ direction.

\begin{figure}[htbp]
\centering
\includegraphics
[scale=0.32]{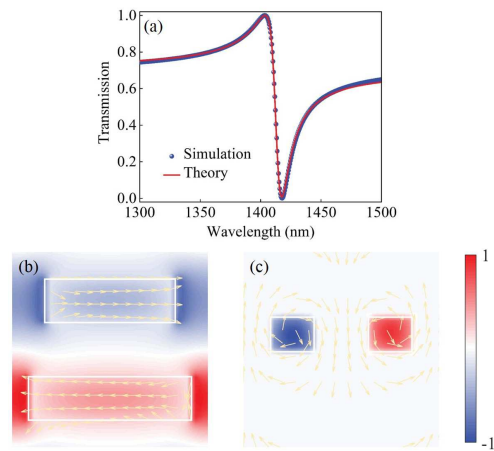}
\caption{\label{fig2} (a) Simulated and fitted transmission spectra of the dielectric metasurfaces. (b) Magnitude of the $x$-component of electric field across the unit cell at the resonance wavelength, overlaid with arrows indicating direction of displacement current. (c) Corresponding magnitude of the $x$-component of displacement current, overlaid with arrows indicating direction of magnetic field.}
\end{figure}

To verify the physical picture of the radiation control at critical coupling via BIC, we first perform the resonance mode analysis without the presence of graphene. The lengths of the nanobars are set with $L_{1}=600$ nm and $L_{2}=750$ nm, respectively. Due to the in-plane symmetry breaking of the unit cell, BIC mode manifests itself in the transmission spectrum as a Fano resonance. As shown in FIG. \ref{fig2}(a), the periodic dielectric structure shows a pronounced asymmetric resonance with a narrow dip at 1417.99 nm, which is well fitted by the classical Fano formula within CMT framework\cite{Hsu2013}. The radiation rate is extracted as $\gamma=5.99$ THz while the dissipative loss rate is zero due to the lossless silicon nanobars here. In FIG. \ref{fig2}(b) and \ref{fig2}(c), the nanobars in the unit cell displays a magnetic dipole mode, where an anti-phase and almost equal amplitude electric polarizations are observed inside the silicon nanobars and the circulating displacement currents in the nanobars give rise to out-of-plane magnetic dipole moments. The field profiles reveal the quasi-BIC magnetic dipole nature of the proposed metasurfaces with a highly confined electromagnetic field inside nanobars, and this trapped mode offers a platform to engineer light-matter interactions.

\begin{figure}[htbp]
\centering
\includegraphics
[scale=0.44]{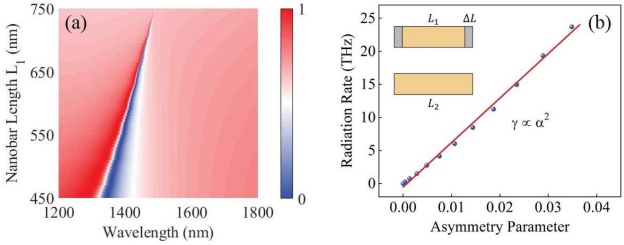}
\caption{\label{fig3} (a) Transmission spectra of the dielectric metasurfaces as a function of the length $L_{1}$ of the shorter nanobar. (b) Values of radiation rate $\gamma$ as a function of the asymmetry parameter $\alpha$. Inset: the asymmetry parameter $\alpha$ is defined as $\Delta L/L_{2}$, where $\Delta L$ is half the difference between nanobar lengths $L_{1}$ and $L_{2}$.}
\end{figure}

Since the quasi-BIC mode is involved with the external radiation channels, the radiation rate and thus the resonance width can be flexibly controlled through manipulating the asymmetry degree. In the proposed dielectric metasurfaces, we observe that transmission spectra show a reducing bandwidth as length of the shorter nanobar $L_{1}$ approaches that of the longer nanobar $L_{2}$ in FIG. \ref{fig3}(a), which can be attributed to the decreasing in the radiation rate $\gamma$ of the dielectric metasurface arising from the reduced asymmetry during the period. For a quantitative description, the asymmetry parameter $\alpha$ here is defined as the ratio of the reduced length $\Delta L$ to the length $L_{2}$ and it can be varied by simply varying $L_{1}$ while fixing $L_{2}$. The values of radiation rate $\gamma$ as a function of the asymmetry parameter $\alpha$ for the dielectric metasurface are summarized in FIG. \ref{fig3}(b). It can be seen clearly that a quadratic dependence of $\gamma$ on $\alpha$ is consistent with the theoretical relation in Eq. (\ref{eq10}) for asymmetric resonances governed by BIC, i.e., $\gamma\propto\alpha^{2}$. Constrained with the general theory for absorption bandwidth in Eq. (\ref{eq11}), the reducing value of $\alpha$ as the length $L_{1}$ increases to $L_{2}$ accounts for the narrowing bandwidth, which lays the foundation for the FWHM manipulation dependent on the asymmetry parameter.

\begin{figure}[htbp]
\centering
\includegraphics
[scale=0.44]{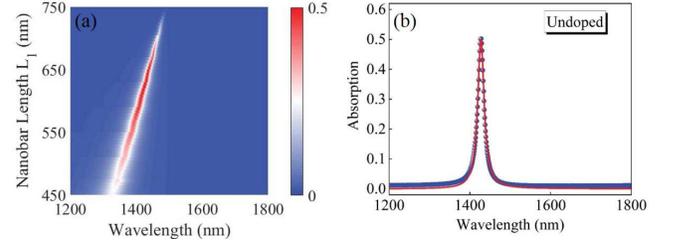}
\caption{\label{fig4} (a) Absorption spectra of the hybrid graphene-dielectric metasurfaces as a function of the length $L_{1}$ of the shorter nanobar. (b) Simulated and fitted absorption spectra for undoped graphene at critical coupling.}
\end{figure}

To construct a critical coupling system, a monolayer graphene is taken into account as a lossy component on top of the silicon nanobars. In the hybrid metasurfaces, the inserted graphene determines the dissipative loss rate $\delta$, while the radiation rate $\gamma$ is dependent on the asymmetry degree of nanobars as above. When these two parameters are equal in the system, i.e., $\gamma=\delta$, the critical coupling condition is satisfied and a maximum absorption is to be achieved. FIG. \ref{fig4}(a) illustrates the absorption spectra for undoped graphene with respect to the nanobar length $L_{1}$ and incident wavelength. The critical coupling point can be found by sweeping these two variables, and the maximum absorption of 0.5 is attained at 1427.39 nm with the nanobar length $L_{1}=625$ nm. The simulated result is further verified by the fitting curve from Eq. (\ref{eq4}), as shown in \ref{fig4}(b). The values of the radiation rate and dissipative loss rate can be extracted as $\gamma=\delta=4.41$ THz, and the corresponding absorption bandwidth is $\Gamma^{\text{FWHM}}=19.06$ nm, showing a good match with the theoretical expression in Eq. (\ref{eq6}).

\begin{figure}[htbp]
\centering
\includegraphics
[scale=0.44]{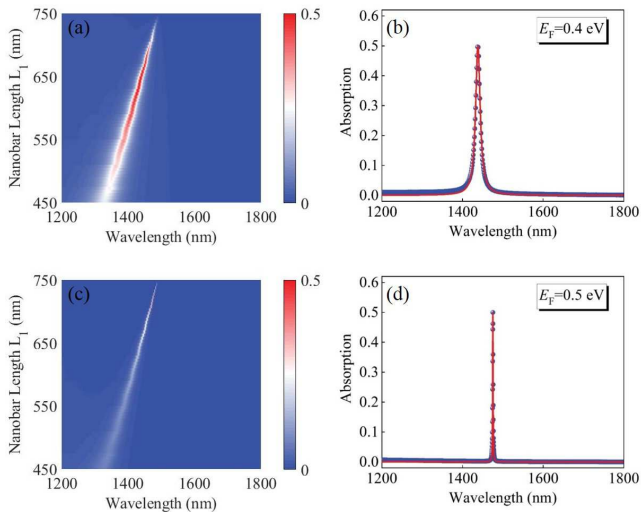}
\caption{\label{fig5} (a, c) Absorption spectra of the hybrid graphene-dielectric metasurfaces as a function of the length $L_{1}$ of the shorter nanobar. (b, d) Simulated and fitted absorption spectra for doped graphene at critical coupling.}
\end{figure}

Next we consider the absorption bandwidth at critical coupling when the monolayer graphene becomes doped. Based on the fact that the radiation rate $\gamma$ and the dissipative loss rate $\delta$ play the crucial roles in determining the absorption characteristics, we control these two parameters to realize such manipulation. More specifically, $\delta$ is changed by tuning graphene conductivity via shifting its Fermi level, while $\gamma$ is varied by adjusting the structural asymmetry of dielectric metasurfaces. The absorption spectra for doped graphene with respect to the nanobar length $L_{1}$ and incident wavelength are plotted in FIG. \ref{fig5}(a) and \ref{fig5}(c), respectively. Here the Fermi levels of 0.4 and 0.5 eV are adapted because the graphene conductivity exhibits a dramatic decrease across the Pauli-blocking point, which will remarkably reduce the dissipative loss. In a similar way with FIG. \ref{fig4}, the critical coupling points can be found by sweeping the nanobar length $L_{1}$ and incident wavelength, and the maximum absorption is attained at 1438.13 nm with $L_{1}=645$ nm for $E_{\text{F}}=0.4$ eV and at 1475.26 nm with $L_{1}=720$ nm for $E_{\text{F}}=0.5$ eV, respectively. The simulated results are once again verified by the fitting curves from Eq. (\ref{eq4}), as shown in FIG. \ref{fig5}(b) and \ref{fig5}(d). Both of the spectra share an absorption peak of 0.5 in agreement with theoretical maximum. The values of the radiation rate and dissipative loss rate can be extracted as $\gamma=\delta=3.11$ THz and $\gamma=\delta=0.27$ THz. As a result of the simultaneous decreasing of $\gamma$ and $\delta$, the absorption bandwidth shrinks by more than one order of magnitude from 13.60 nm to 1.21 nm.

\begin{figure}[htbp]
\centering
\includegraphics
[scale=0.32]{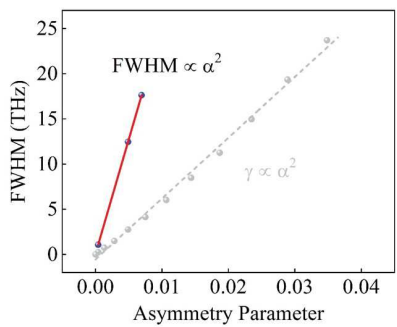}
\caption{\label{fig6} Quadratic dependence of the absorption bandwidth FWHM (red line) and the radiation rate $\gamma$ (grey line) on the asymmetry parameter $\alpha$.}
\end{figure}

We further provide the dependence of absorption bandwidth FWHM on the asymmetry parameter $\alpha$ with the different Fermi levels of graphene. Extracted from the critical coupling condition corresponding to the cases undoped graphene and doped graphene with 0.4 eV and 0.5 eV, the estimated FWHM is proportional to the square of $\alpha$ as can be seen from FIG. \ref{fig6}, i.e., $\Gamma^{\text{FWHM}}\propto\alpha^{2}$, and even more interesting, the slope of this linear fitting is four times that of the radiation rate in FIG. \ref{fig3}(b), which confirms the analytical description in Eq. (\ref{eq11}). It is noted that the quadratic relation between the radiation rate $\gamma$ with the asymmetric parameter $\alpha$ remains valid for the general case of asymmetry metasurfaces with high-quality resonance governed by BIC, and by parity of reasoning the quadratic scalability of FWHM at critical coupling shows general applicability, offering a straightforward way to engineer the absorption properties.

\section{\label{sec4}Conclusions}
In conclusions, we bridge a direct link between the radiation control at critical coupling and the intriguing BIC physics, and develop a generalized theory to engineer light absorption of 2D materials in coupling resonance systems. The quadratic dependence of the absorption bandwidth on the asymmetry parameter is obtained. In the hybrid graphene-dielectric metasurfaces, the absorption bandwidth can be flexibly tuned by more than one order of magnitude by simultaneously adjusting the asymmetry parameter for the silicon nanobar resonators and the graphene surface conductivity while the absorption efficiency maintains maximum. Beyond this typical example, the proposed theoretical method should have wide applicability to diverse types of critical coupling systems, in principle, with various asymmetric metasurfaces designs and different atomically thin 2D materials throughout the spectrum. Therefore, this work not only gains a deeper physical insight into the enhanced light-matter interaction in 2D materials, but also paves a way towards smart design of compact optoelectronic devices such as light emitters, detectors and modulators.

\begin{acknowledgments}	
This work is supported by National Natural Science Foundation of China (Grant No. 11664024, 11847132, 11947065 and 61901164), Key Project of Natural Science Foundation of Jiangxi Province (Grant No. 20171ACB201020), Interdisciplinary Innovation Fund of Nanchang University (Grant No. 2019-9166-27060003), Science and Technology Foundation of Guizhou Province \& Guizhou Minzu University (Grant No. LKM[2012]25), Natural Science Research Project of Guizhou Minzu University (Grant No. GZMU[2019]YB22) and China Scholarship Council (Grant No. 202008420045).
\end{acknowledgments}

\end{document}